\documentclass[aps,prl,twocolumn,showpacs,superscriptaddress,floatfix,nofootinbib]{revtex4}
\usepackage{graphicx}
\usepackage{amsmath}
\usepackage{epsfig}
\usepackage{helvet}
\usepackage{amssymb}

\newcommand{\be}{\begin{equation}}
\newcommand{\ee}{\end{equation}}
\newcommand{\bea}{\begin{eqnarray}}
\newcommand{\eea}{\end{eqnarray}}

\newcommand{\bra}[1]{\mbox{$\langle #1 |$}}
\newcommand{\ket}[1]{\mbox{$| #1 \rangle$}}
\newcommand{\braket}[2]{\mbox{$\langle #1  | #2 \rangle$}}
\newcommand{\proj}[1]{\mbox{$|#1\rangle \!\langle #1 |$}}

\begin{document}

\title{
Classical simulation of infinite-size quantum lattice systems in one
spatial dimension}
\author{G. Vidal}
\affiliation{School of Physical Sciences, the University of
Queensland, QLD 4072, Australia}

\date{\today}

\begin{abstract}
Invariance under translation is exploited to efficiently
simulate one-dimensional quantum lattice systems in the limit of an
infinite lattice. Both the computation of the ground state and the
simulation of time evolution are considered.
\end{abstract}


\maketitle


Numerical renormalization group (RG) methods obtained a first
remarkable success with Wilson's solution of the Kondo problem
\cite{wilson} and, ever since the irruption of White's density
matrix renormalization group (DMRG) algorithm \cite{white}, are
solidly established as the dominant computational approach to
quantum lattice systems in one spatial dimension (1D)
\cite{schollwoeck}. Recently, progress in our understanding of
quantum entanglement has prompted a series of new developments
\cite{tebd,peps,frank_others,zwolak,tobias,tree} that extend the
domain of RG methods in two main directions. On the one hand,
still in the context of 1D lattice systems, the {\em time-evolving
block decimation algorithm} (TEBD) allows for the simulation of time
evolution \cite{tebd}. On the other, Verstraete and Cirac
\cite{peps} have introduced {\em projected entangled-pair states}
(PEPS) to address the simulation of quantum lattice systems in two
and higher spatial dimensions.


In this work we propose an algorithm, based on TEBD, to simulate 1D quantum
lattice systems in the thermodynamic limit. Bulk properties of
matter are best studied in an infinite system, where they are not
contaminated by finite-size corrections or boundary effects.
However, for most algorithms the cost of a simulation grows with the
system size, and the thermodynamic limit can only be reached by
extrapolating results for increasingly large systems. 
Here we exploit two facts, namely {\em invariance under translations} of the system and {\em parallelizability} of local updates in TEBD, to obtain a noticeably simple and fast algorithm, referred to as {\em infinite} TEBD or iTEBD, to
simulate infinite systems directly, without resorting
to costly, less accurate extrapolations. 
We describe the iTEBD algorithm and
test its performance by computing the ground state and time evolution for a quantum
spin chain in the thermodynamic limit. In addition to offering a
very competitive alternative to DMRG for infinite systems, iTEBD plays a key role in {\em entanglement renormalization} techniques
\cite{ent_ren} and in the extension of PEPS \cite{peps} to infinite 2D lattices \cite{ipeps}.


We consider an infinite array of sites in 1D, where each site $r$,
$r\in\mathbb{Z}$, is described by a complex vector space
$V^{[r]}\cong \mathbb{C}^d$ of finite dimension $d$. Let vector
$\ket{\Psi}$,
\begin{equation}\label{eq:Psi}
    \ket{\Psi} \in \bigotimes_{r\in \mathbb{Z}}
    V^{[r]},
\end{equation}
denote a pure state of the lattice and operator $H$,
\begin{equation}\label{eq:ham}
    H = \sum_{r} h^{[r,r+1]},
\end{equation}
a Hamiltonian with nearest neighbor interactions, and let 
$\ket{\Psi}$ and $H$ be invariant under shifts by one lattice site \cite{extend}. 
Given an initial state
$\ket{\Psi_0}$, our goal is to simulate an evolution
according to $H$, both in real time
\begin{equation}\label{eq:evolution}
    \ket{\Psi_t} = \exp(-i Ht)\ket{\Psi_0},
\end{equation}
and in imaginary time \cite{infinite},
\begin{equation}\label{eq:evolution2}
    \ket{\Psi_\tau} =
    \frac{\exp(-H\tau)\ket{\Psi_0}}{\|\exp(-H\tau)\ket{\Psi_0}\|}.
\end{equation}


The TEBD algorithm represents $\ket{\Psi}$ through a matrix product
state (MPS) \cite{mps}. See \cite{tebd} for details on this
structure, that we briefly review for an infinite 1D lattice.
Let $[\lhd r]$ and $[r+\!1 \rhd]$ denote the semi-infinite
sublattices made of sites $\{-\infty,\cdots, r\}$ and
$\{r+\!1,\cdots, \infty\}$. If the Schmidt decomposition
of $\ket{\Psi}$ according to this bipartition reads
\begin{equation}\label{eq:Schmidt}
    \ket{\Psi} = \sum_{\alpha=1}^{\chi} \lambda_{\alpha}^{[r]}
    \ket{\Phi^{[\lhd r]}_{\alpha}}\otimes\ket{\Phi^{[r+\!1
    \rhd]}_{\alpha}},
\end{equation}
where we assume the Schmidt rank $\chi$ to be finite, then the
spectral decomposition of the reduced density matrices for $[\lhd r]$ and $[r+\!1 \rhd]$ are
\begin{eqnarray}\label{eq:reduced}
 \rho^{[\lhd r]} &=&
\sum_{\alpha=1}^{\chi}~(\lambda_{\alpha}^{[r]})^2~
\proj{\Phi^{[\lhd r]}_{\alpha}}, \\
 \rho^{[r+\!1 \rhd]} &=&
\sum_{\alpha=1}^{\chi}~(\lambda_{\alpha}^{[r]})^2~
\proj{\Phi^{[r+\!1\rhd]}_{\alpha}}.
\end{eqnarray}
We use a three-index tensor $\Gamma^{[r]}$ to relate the Schmidt
bases for two left (respectively right) sublattices:
\begin{eqnarray}\label{eq:Gamma}
\ket{\Phi^{[\lhd r+1]}_{\alpha}} &=& \sum_{\beta=1}^\chi\sum_{i=1}^d
\lambda^{[r]}_{\beta}\Gamma^{[r\!+1]}_{i\beta\alpha}
~\ket{\Phi^{[\lhd
r]}_{\beta}} \ket{i^{[r\!+1]}}, \\
 \ket{\Phi^{[r\rhd]}_{\alpha}} &=& \sum_{\beta=1}^\chi\sum_{i=1}^d
\Gamma^{[r\!+1]}_{i\alpha\beta}
\lambda^{[r\!+1]}_{\beta}~\ket{i^{[r]}} \ket{\Phi^{[r+1\rhd
]}_{\beta}}.
\end{eqnarray}
\begin{figure}
  \includegraphics[width=8cm]{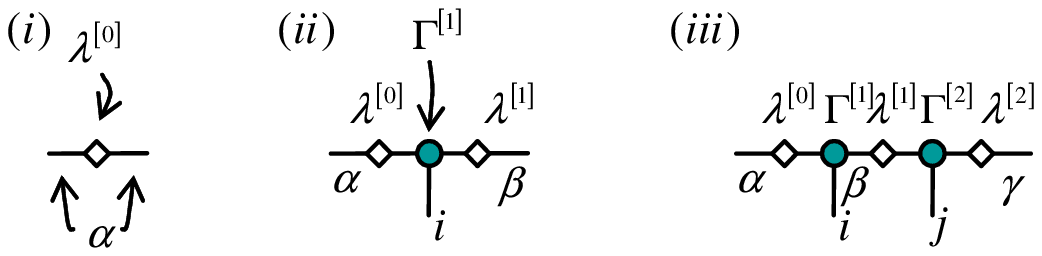}\\
\caption{Diagrammatic representation of several decompositions of
$\ket{\Psi}$ in terms of a network of tensors, see \cite{tree} for
more details. Each filled circle represents a tensor $\Gamma$ and
each edge represents an index, such as $\alpha$ or $i$. Hanging from
an open side of an edge there is an (omitted) orthonormal basis,
such as $\ket{\Phi^{[\lhd 0]}_{\alpha}}$ or $\ket{i^{[1]}}$. An
empty diamond on top of an edge represents a set of weights
$\lambda$ (Schmidt coefficients) for the corresponding index. ({\em
i}) Schmidt decomposition, Eq. (\ref{eq:Schmidt}), according to
semi-infinite sublattices $[\lhd 0]$ and $[1 \rhd]$. The bases
$\ket{\Phi^{[\lhd 0]}_{\alpha}}$ and $\ket{\Phi^{[1
\rhd]}_{\alpha}}$ are linked through index $\alpha$ with weights
$\lambda^{[0]}_{\alpha}$. ({\em ii}) Expansion of $\ket{\Psi}$ in
terms of the Schmidt bases for the semi-infinite sublattices $[\lhd
0]$ and $[2 \rhd]$ and an orthonormal basis for site $1$, see Eq.
(\ref{eq:1site}). ({\em iii}) Expansion of $\ket{\Psi}$ in terms of
orthonormal sets of vectors for sublattices $[\lhd 0]$ and $[3
\rhd]$ and sites $1$ and $2$.}\label{fig:Schmidt}
\end{figure}
In particular, $\ket{\Psi}$ can be expanded in the local basis $\ket{i^{[r]}}$ for
site $r$ and in terms of
$\lambda^{[r]}\Gamma^{[r\!+1]}\lambda^{[r\!+1]}$ as
\begin{equation}\label{eq:1site}
    \ket{\Psi} \!= \!\sum_{\alpha,\beta=1}^{\chi} \sum_{i=1}^d
    \lambda_{\alpha}^{[r]}\Gamma^{[r\!+1]}_{i\alpha\beta}
\lambda^{[r\!+1]}_{\beta}
    \ket{\Phi^{[\lhd r]}_{\alpha}}\ket{i^{[r]}}\ket{\Phi^{[r+\!1
    \rhd]}_{\alpha}},
\end{equation}
or for sites $\{r, r+1\}$ in terms of
$\lambda^{[r]}\Gamma^{[r\!+1]}\lambda^{[r\!+1]}\Gamma^{[r\!+2]}\lambda^{[r\!+2]}$,
and so on, see Fig.(\ref{fig:Schmidt}).
We also recall that in the TEBD algorithm the evolution operator
$\exp(-iHt)$ in Eq. (\ref{eq:evolution}) is expanded through a
Suzuki-Trotter decomposition \cite{trotter} as a sequence of small
two-site gates
\begin{equation}\label{eq:gate}
    U^{[r,r+1]} \equiv \exp(-ih^{[r,r+1]}\delta t),  ~~~~~~ \delta t \ll
    1,
\end{equation}
which we arrange into gates $U^{AB}$ and $U^{BA}$,
\begin{eqnarray}\label{eq:suzuki-trotter}
U^{AB}\equiv \bigotimes_{r\in\mathbb{Z}}
U^{[2r,2r\!+1]},~~~~~~~U^{BA}\equiv \bigotimes_{r\in\mathbb{Z}}
U^{[2r\!-1,2r]}.
\end{eqnarray}

\begin{figure}
  \includegraphics[width=6cm]{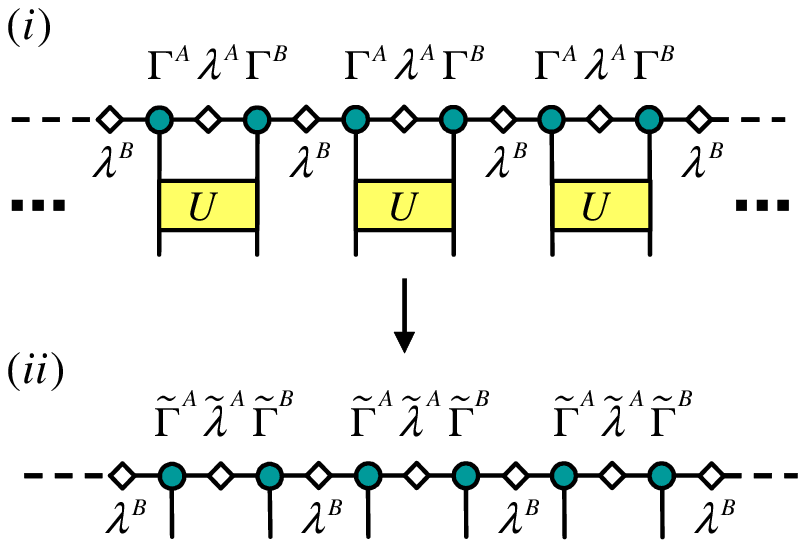}\\
\caption{ Two diagrammatical representations (see also Fig.
(\ref{fig:Schmidt}) and ref. \cite{tree}) for state
$U^{AB}\ket{\Psi}$: ({\em i}) tensor network representation
containing both an MPS for $\ket{\Psi}$ and two-site gates $U$
acting on each pair of sites $\{2r, 2r+1\}$, $\forall
r\in\mathbb{Z}$; ({\em ii}) new MPS for $U^{AB} \ket{\Psi}$. Notice
that both structures are invariant under shifts by two lattice sites
and are completely specified by a small number of tensors, in spite
of the fact that they represent a state of an infinite 1D
lattice.}\label{fig:evolution}
\end{figure}
\begin{figure}
  \includegraphics[width=8cm]{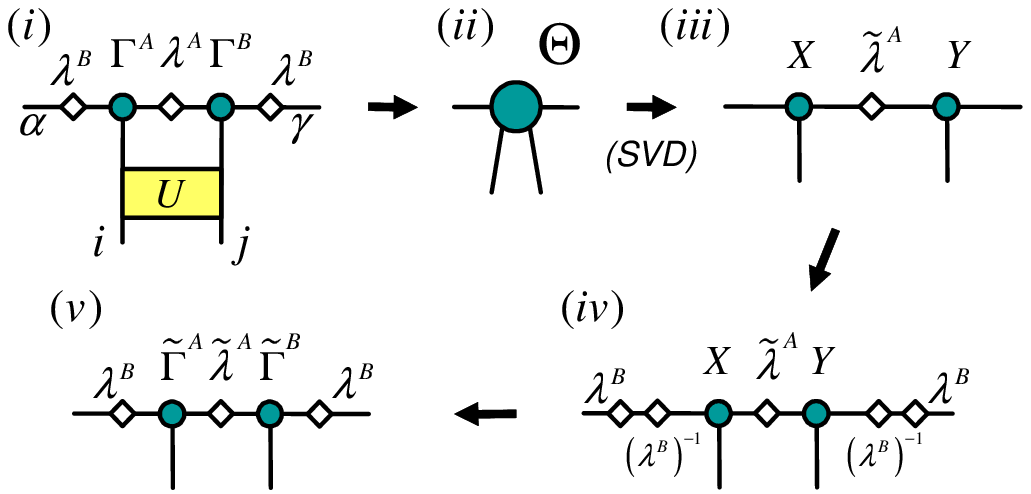}\\
\caption{In order to update the MPS after gate $U$ has been applied,
see Fig. (\ref{fig:evolution}), we first contract the tensor network
({\em i}) into a single tensor $\Theta_{\alpha i j \gamma}$ ({\em
ii}). We then compute the {\em singular value decomposition} of
$\Theta$ according to the index bipartition $[\alpha i]:[j\gamma]$,
namely $\Theta = \sum_{\beta} X_{[\alpha
i]\beta}\tilde{\lambda}^{A}_\beta Y_{\beta[j \gamma]}$ as in ({\em
iii}). We introduce $\lambda^B$ back into the network ({\em iv}) and
form tensors $\tilde{\Gamma}^{A}$ and $\tilde{\Gamma}^{[B]}$ in
({\em iv}) by attaching to $X$ and $Y$ the inverse of the Schmidt
coefficients $\lambda^{B}$. All such matrix manipulations are
achieved with $O(d^2\chi^2)$ space and $O(d^3\chi^3)$ and need to be
performed only once in order to update the MPS for the whole
infinite chain.}\label{fig:svd}
\end{figure}

Because state $\ket{\Psi}$ is shift invariant, it could be represented
with a MPS where $\Gamma^{[r]}$ and $\lambda^{[r]}$ are independent
of $r$. However, we will partially break
translational symmetry to simulate the action of gates
(\ref{eq:suzuki-trotter}) on $\ket{\Psi}$. Accordingly, we choose a
MPS of the form
\begin{eqnarray}\label{eq:GammaAB}
    \Gamma^{[2r]} &=& \Gamma^{A}, ~~~~~~~\lambda^{[2r]} = \lambda^A,\nonumber \\
    \Gamma^{[2r\!+1]} &=& \Gamma^{B}, ~~~~\lambda^{[2r\!+1]} =
    \lambda^B,~~~r\in\mathbb{Z}.
\end{eqnarray}
As in the finite $n$ case \cite{tebd}, the simulation of the time evolution, see Eq.
(\ref{eq:evolution}), is achieved by updating the MPS so as to
account for the repeated application of gates $U^{AB}$ and $U^{BA}$
on $\ket{\Psi}$. But for $n = \infty$, the action of the
gates preserve the invariance of the evolved state under shifts by
{\em two} sites, see Fig. (\ref{fig:evolution}), and only tensors
$\Gamma^{A}, \Gamma^{B},\lambda^{A}$ and $\lambda^{B}$ need to be
updated -- a task that is achieved through simple matrix
manipulations, see Fig. (\ref{fig:svd}). In other words, whereas for $n$ sites the TEBD algorithm requires
$O(nd\chi^2)$ space to store an MPS and $O(nd^3\chi^3)$ time to
simulate a small evolution $\exp(-iH\delta t)$ \cite{tebd}, for $n = \infty$ sites the iTEBD requires computational
space and time that scale just as $O(d^2\chi^2)$ and $O(d^3\chi^3)$.
Key to such dramatic cost reduction by a factor $n$ is the fact
that, in contrast to other approaches \cite{frank_others}, we use a
MPS based on the Schmidt decomposition, allowing for a parallelized, {\em local}
update of tensors $\Gamma$ and $\lambda$. 

Finally, evolution in imaginary time, Eq. (\ref{eq:evolution2}), is
also simulated with iTEBD by simply replacing the two-site
unitary gates $\exp(-i h~\delta t)$ in Eq. (\ref{eq:gate}) with
non-unitary gates $\exp(-h ~\delta\tau)$, $\delta \tau \ll 1$
\cite{ignore}.

\begin{figure}
  \includegraphics[width=9.5cm]{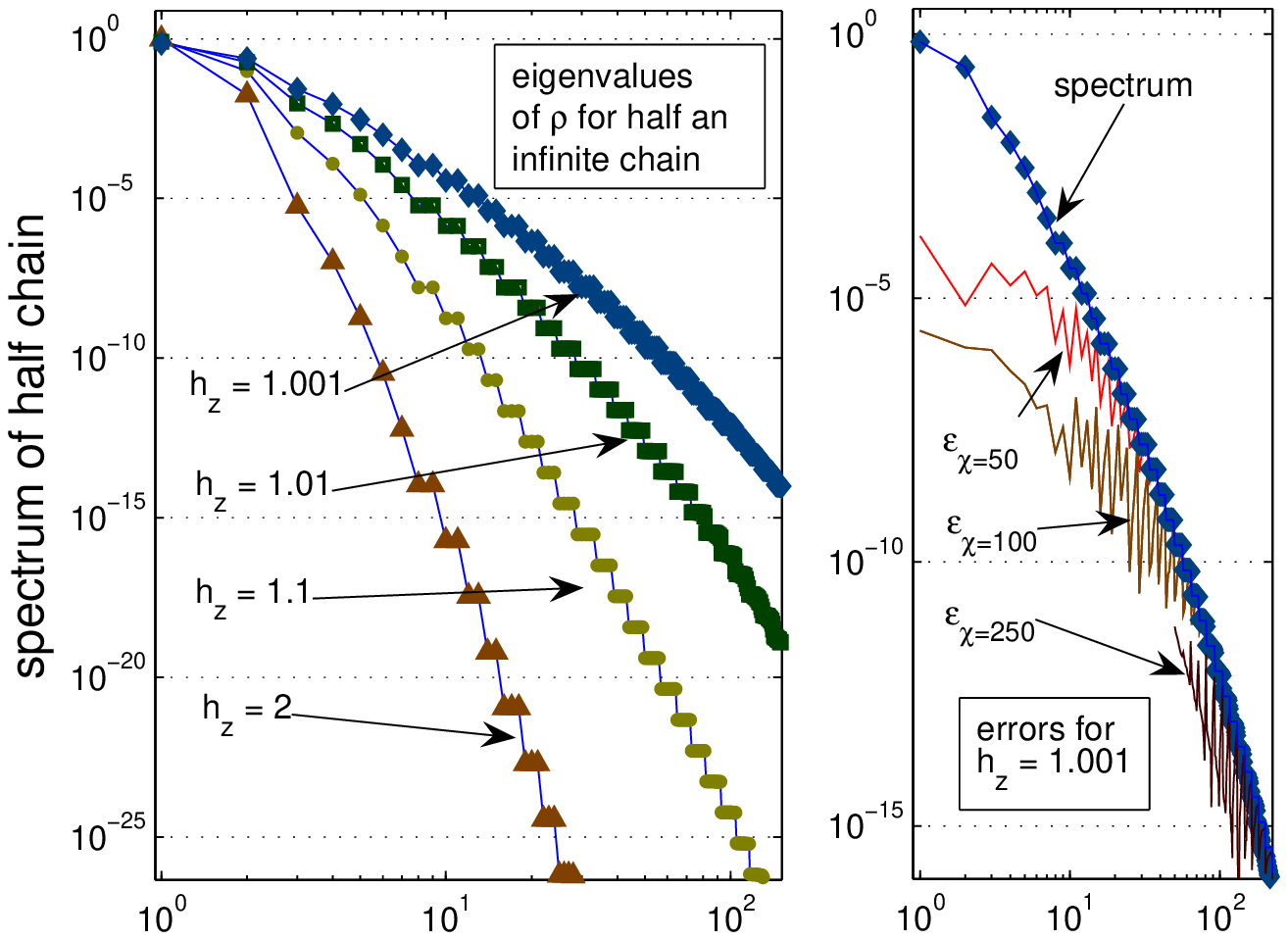}\\
  \caption{
Left: spectrum $p_{\alpha} = (\lambda_{\alpha})^2$ of the density
matrix for half of the infinite chain, see Eq. (\ref{eq:reduced}),
for different values of the magnetic field $h_z$. For $h_z \gg 1$
the eigenvalues $p_{\alpha}$ decay very fast with $\alpha$. As $h_z$
approaches the critical value $h_z^*=1$, the decay is less
pronounced and the computation of the spectrum becomes more
demanding. Right: however, for $h_z=1.001$ we still obtain the 50,
100 and 250 first eigenvalues with remarkable accuracy, the error on
$p_{\alpha}$ being typically several orders of magnitude smaller
that its value. }\label{fig:spectrum}
\end{figure}

We have tested the performance of iTEBD by computing the
ground state and by simulating time evolution for the 1D quantum
Ising chain with transverse magnetic field, as defined by the
Hamiltonian
\begin{equation}\label{eq:Ising}
H = \sum_{r\in \mathbb{Z}} (\sigma_x^{[r]}\sigma_x^{[r\!+\!1]} +
h_{z}
\sigma_z^{[r]}).
\end{equation}
This model is exactly solvable \cite{ising_exact}, allowing for a
direct comparison of numerical results with the exact solution.

\begin{figure}
  \includegraphics[width=9.5cm]{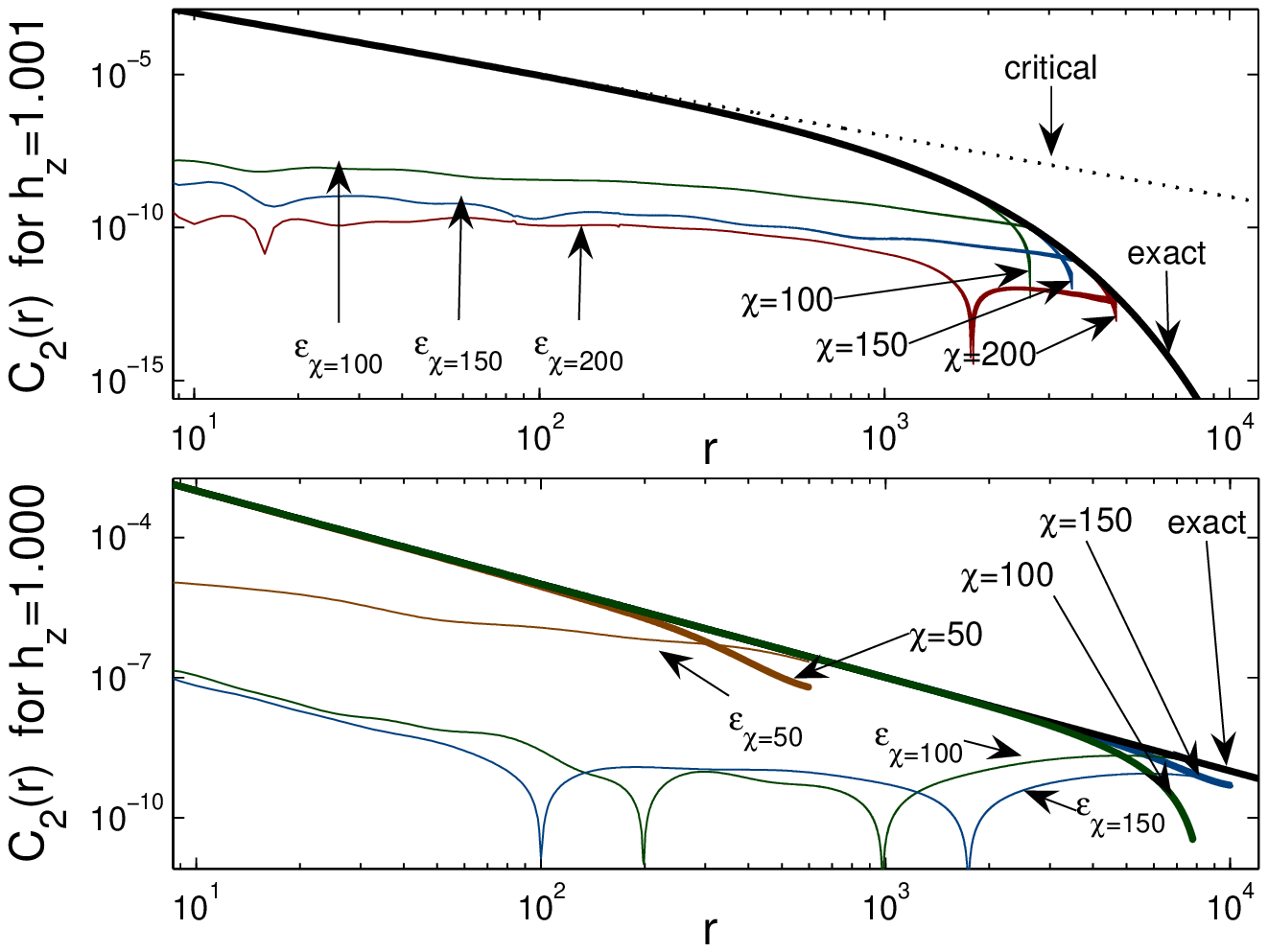}\\
  \caption{
Two point correlator $C_2(r)$ of Eq. (\ref{eq:zz}). Up: for $h_z =
1.001$, when the distance $r$ between spins is up to several hundred
sites, $C_2(r)$ seems to decay as a power law (as in the critical
case) but for $r$ of the order of $1000$ the exponential character
of the decay becomes manifest. Down: even though iTEBD is unable to properly reproduce the spectrum $p_{\alpha}$
of half an infinite chain at criticality, $h_z=1$, (see Fig.
(\ref{fig:spectrum})), we still obtain a very accurate approximation
to the correlator $C_2(r)$ when $r$ is of the order of several
thousands of spins.
  }\label{fig:zz}
\end{figure}

Firstly, by simulating an evolution in imaginary time, Eq.
(\ref{eq:evolution2}), we have obtained an approximate MPS representation of the
ground state $\ket{\Psi_{g}}$ for several values of $h_z \geq 1$. We
assess the accuracy of the result by focussing on the
following: ({\em i}) the spectrum $p_{\alpha} = (\lambda^{[r]}_{\alpha})^2$ of the reduced
density matrix for half the infinite lattice, Eq.
(\ref{eq:reduced}), which
characterizes entanglement accross the boundary of two halves of the
chain; and ({\em ii}) the two-point correlator
\begin{equation}\label{eq:zz}
C_2(r)  \equiv  \bra{\Psi_{g}} \sigma_z^{[0]} \sigma_z^{[r]}
\ket{\Psi_{g}} - (\bra{\Psi_{g}} \sigma_z^{[0]} \ket{\Psi_{g}})^2,
\end{equation}
which quantifies the strength of correlations between two spins that
are $r$ lattice sites apart.

In both cases, we obtain quantitative agreement with the exact
solution up to several significant digits, see Figs.
(\ref{fig:spectrum}) and (\ref{fig:zz}). Computing the ground state
$\ket{\Psi_g}$ is particularly fast and accurate for large values of
$h_z$ (say $h_z\geq 1.1$) and, as expected, becomes more resource intensive as
$h_z$ approaches the critical value $h_z^*=1$ \cite{resources}.
We find, however, that for
values very close to the critical point, such as $h_z=1.001$, we
still obtain accurate approximations to the half-chain spectrum
$\{p_{\alpha}\}$ and the correlator $C_2(r)$ through affordable
simulations, see Fig. (\ref{fig:zz}). And, most remarkably, reliable
results for $C_2(r)$ are obtained even at the critical point $h_z=1$
by using a MPS with a reasonably small $\chi$, in spite of the fact
that such a MPS is no longer able to reproduce $\{p_{\alpha}\}$.

\begin{figure}
  \includegraphics[width=9.5cm]{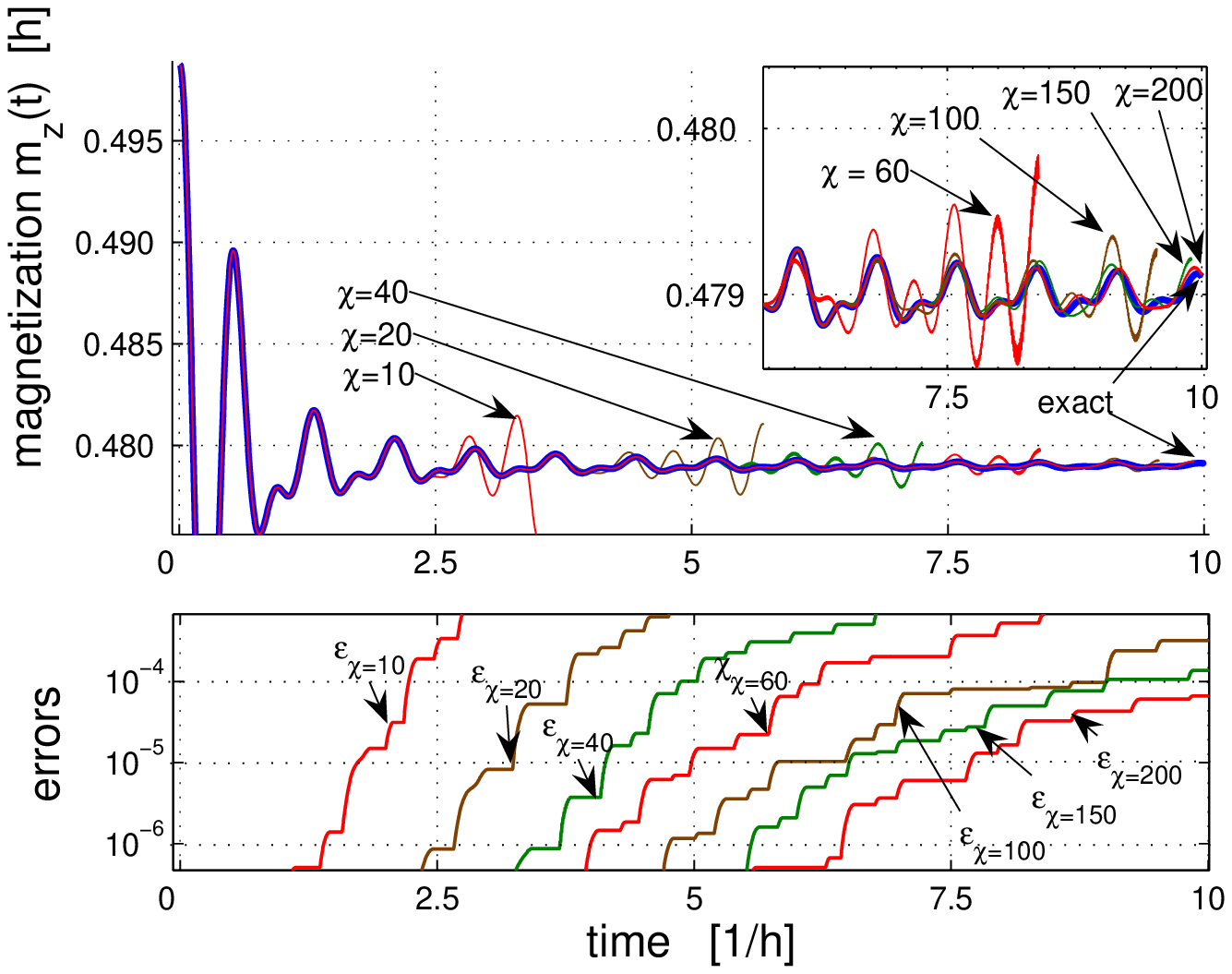}\\
  \caption{
Simulation of the evolution in time of $m_z(t)$ in Eq.
(\ref{eq:magnet}). The infinite chain is initially in the ground
state of $H$ with $h_z=10$. At $t=0$ the magnetic field is changed
to $h_z=3$. Up: the magnetization, originally at a value
$m_z(0)\approx 0.5$, oscillates in time until it stabilizes around a
lower value $m_z(\infty) \approx 0.48$. By considering increasingly
large values of $\chi$ we are able to simulate the evolution of
$m_z(t)$ for long times with remarkable accuracy.
  }\label{fig:time_evolution}
\end{figure}

Secondly, by switching to real time, we have simulated the
evolution of the infinite system, prepared in the ground state of
$H$ in Eq. (\ref{eq:Ising}) for $h_z = 10$, when the magnetic field
is abruptly modified from its initial value to $h_z'=3$. Fig.
(\ref{fig:time_evolution}) shows the evolution in time of the
magnetization
\begin{equation}\label{eq:magnet}
    m_z(t) \equiv \bra{\Psi_{g}} \sigma_z^{[1]}
\ket{\Psi_{g}},
\end{equation}
which after several oscillations becomes stable at some value
different from the initial one.

Summarizing, with modest computational resources the iTEBD is able
to analyze an infinite 1D system, near to and at a quantum critical
point, for which it obtained accurate results for quantities
involving up to several thousands of sites. Comparable results with
previous approaches \cite{white,tebd,frank_others} would require
computations that are, to the best of our knowledge,
unaffordable. Indeed, simulating a 1D lattice of, say, $n=10,000$
sites, would have a time and memory cost proportional to $n$, and would
still include $O(1/n)$ errors due to the finite size of the system, in
addition to more important deviations near the boundaries of the
lattice.

Work in progress considers a range of
applications of the iTEBD, including the characterization of quantum critical
points and cross-over between quantum phases, and the study
of the response of 1D systems to external probes. 
Originally developed to help in the
context of entanglement renormalization \cite{ent_ren}, the method
is also the key to extend 2D PEPS \cite{peps} to infinite systems \cite{ipeps}.

The author acknowledges support from the Australian Research Council
through a Federation Fellowship.


\begin{thebibliography}{99}
\bibitem{infinite} For a finite lattice with $n$ sites and a hamiltonian $H$ with gap $\Delta > 0$, simulating imaginary time evolution as in Eq. (\ref{eq:evolution2}) for large $\tau$ yields a good approximation $\ket{\Psi_{\tau}}$ to the ground state 
$\ket{\Psi_{g}}$ of $H$, since one can show that 
\begin{equation}\label{eq:overlap}
    |\braket{\Psi_\tau}{\Psi_{g}}| > 1 -
    O(\frac{e^{-2\Delta\tau}}{\delta^2}), ~~~~\delta \equiv
    |\braket{\Psi_0}{\Psi_{g}}|.
\end{equation}
Notice that Eq. (\ref{eq:overlap}) may becomes meaningless in the limit $n\rightarrow \infty$, where typically $\delta\rightarrow 0$, or for a gapless hamiltonian, $\Delta = 0$. Here we will not provide a justification for the validity of TEBD, as a method to approximate $\ket{\Psi_g}$, in the $n\rightarrow \infty$ and/or $\Delta = 0$ regimes, but refer instead to successful benchmark calculations.
\bibitem{ignore}
Non-unitary gates destroy the Schmidt decomposition and a parallelized, local update of tensors $(\ref{eq:GammaAB})$ should no
longer be possible. For small $\delta \tau$, however, gate $\exp
(-h\delta \tau)$ is close to the identity, and proceeding with a
local update introduces only small errors that vanish as $\delta
\tau \rightarrow 0$. An accurate MPS for the ground state
$\ket{\Psi_g}$ is then achieved by simulating imaginary time
evolution with decreasingly small values of $\delta \tau$.
\bibitem{resources} The simulations were performed using matlab code in a
laptop with a 2 GHz processor and 1 GB of RAM. An approximation to
the ground state of $H$ in Eq. (\ref{eq:Ising}) was computed through
an evolution in imaginary time until convergence of the spectrum
$\{p_{\alpha}\}$. We used a fourth order Suzuki-Trotter expansion
\cite{trotter}, with decreasing values of $\delta \tau \in \{0.1,
0.01, 0.001, \cdots\}$. For $h_z=1.1$ with $\chi=50$, convergence of
the spectrum was achieved after just a few seconds, whereas the case
$h_z=1.01$ with $\chi=100$ required a few minutes and the case
$h_z=1.001$ with $\chi=200$ required a few hours. At criticality,
$h_z=1$ and $\chi=150$, no convergence of the spectrum was achieved
after a few days. However, just after a few hours, the resulting MPS
was able to accurately reproduce the correlator $C_2(r)$ for $r\leq
5000$ sites.
\end{thebibliography}
\end{document}